\documentclass[%
 reprint,
 superscriptaddress,
 amsmath,amssymb,
 aps,
 prb,
]{revtex4-2}

\usepackage{graphicx}
\usepackage{dcolumn}
\usepackage{bm}
\usepackage{hyperref}

\begin{document}

\preprint{APS/123-QED}

\title{Tailoring Dynamical Quantum Phase Transitions via Double-Mode Squeezing Manipulation}

\author{Kaiyuan Cao}
 \email{kycao@yzu.edu.cn}
\affiliation{College of Physics Science and Technology, Yangzhou University, Yangzhou 225009, People's Republic of China}

\author{Haodong Wang}
\affiliation{College of Physics Science and Technology, Yangzhou University, Yangzhou 225009, People's Republic of China}

\author{Xiang-Ping Jiang}
\affiliation{School of Physics, Hangzhou Normal University, Hangzhou, Zhejiang 311121, China}

\author{Shu Chen}
 \email{schen@iphy.ac.cn}
\affiliation{Beijing National Laboratory for Condensed Matter Physics, Institute of Physics, Chinese Academy of Sciences, Beijing 100190, China}

\author{Jian Wang}
 \email{phcwj@hotmail.com}
\affiliation{College of Physics Science and Technology, Yangzhou University, Yangzhou 225009, People's Republic of China}

\date{\today}

\begin{abstract}
    We propose a protocol to tailor dynamical quantum phase transitions (DQPTs) by double-mode squeezing onto the initial state in the XY chain. The effect of squeezing depends critically on the system's symmetry and parameters. When the squeezing operator breaks particle-hole symmetry (PHS), DQPTs become highly tunable, allowing one to either induce transitions within a single phase or suppress them. Remarkably, when PHS is preserved and the squeezing strength reaches $r=\pi/4$, a universal class of DQPTs emerges, independent of the quench path.  This universality is characterized by two key features: (i) the collapse of all Fisher zeros onto the real-time axis, and (ii) the saturation of intermode entanglement to its maximum in each $(k,-k)$ modes. Moreover, the critical momenta governing the DQPTs coincide exactly with the modes attaining the maximal entanglement. At this universal point, the dynamical phase vanishes, leading to a purely geometric evolution marked by $\pi$-jumps in the Pancharatnam geometric phase. Our work establishes initial-state squeezing as a versatile tool for tailoring far-from-equilibrium criticality and reveals a direct link between entanglement saturation and universal nonanalytic dynamics.
\end{abstract}

\maketitle


\textit{Introduction.}---
With the rapid advancement of quantum technologies, precise control over quantum systems has enabled transformative progress across diverse domains, from quantum computation to ultra-sensitive metrology \cite{Qvarfort2018nc, Simon2022SciAdv}. A cornerstone of these developments is the squeezing operator, a fundamental tool that not only lies at the heart of quantum metrology \cite{Taylor2016PhysRep, Taylor2013NatPhoto, Morris2015nc, Marciniak2022nature} but also plays an increasingly vital role in the study of quantum many-body dynamics \cite{Liu2025PRL, Zou2025PRL}. In quantum precision measurement, squeezing is indispensable for enhancing the sensitivity of devices such as atomic clocks \cite{Anders1999PRL, Andre2004PRL, Pezze2020PRL, Bhattacharyya2024PRA, Finkelstein2024nature}, Ramsey spectrometers \cite{Agarwal1996PRA, Xu1999PRA, Sanchez2021PRL, Kenan2013OptLett}, and gravitational-wave detectors \cite{Walls1981PhysLettA, Goda2008NatPhys, Wang2022SciRep}. Beyond metrology, squeezing operations offer a powerful means to steer entanglement, correlations, and collective behavior in interacting quantum systems \cite{Wang2003PRA, Sorensen2001PRL, Tavakoli2024RMP, Cai2025PhysRep, Liu2025PRX, Mazza2025PRB}. Yet, fully harnessing this potential hinges on a central challenge: how to design physically realizable squeezing protocols that are tailored to specific quantum tasks \cite{Ma2011PhysRep}. Addressing this question bridges fundamental theory with practical implementation, representing a critical frontier in the ongoing development of quantum science and technology.

While the applications of squeezing in metrology and equilibrium many-body systems are well recognized, the true potential of quantum control extends into the realm of far-from-equilibrium dynamics \cite{Polkovnikov2011RMP}. In this context, dynamical quantum phase transitions (DQPTs) have emerged as a powerful framework to characterize nonequilibrium critical behavior during the temporal evolution of quantum systems \cite{Zvyagin2016LowTem, Heyl2018RepPro, Heyl2019EuLett}. Unlike conventional phase transitions, which are governed by ground-state properties, DQPTs manifest as nonanalyticities in the Loschmidt echo—a dynamical analogue of the partition function—following a sudden quantum quench \cite{Heyl2013PRL}. These transitions reveal fundamental changes in the dynamical properties of quantum states and offer deep insights into topics such as work statistics \cite{Rylands2019PRB, Campbell2016PRB, Marino2014PRB}, dynamical topological phenomena \cite{Budich2016PRB, Tang2025PRB}, and nonequilibrium scaling laws \cite{Heyl2015PRL, Lupo2016PRB, Zamani2024JPCM, Karrasch2017PRB}. Despite significant theoretical \cite{Jing2024PRL, Vajna2014prb, Schmitt2015prb, Karrasch2013prb, Kriel2014prb, Sharma2015prb, Halimeh2017prb, Homrighausen2017prb, Obuchi2017prb, Halimeh2020prr, Zhou2018pra, Zhou2021prb, Mondal2022prb, Mondal2023prb, Zeng2023prb} and experimental \cite{Vogel2017naturep, Jurcevic2017prl, Chen2020pra, Muniz2020nature, Nie2020prl, Wang2019prl, Xu20209lightsa, Tian2020prl} progress, most studies so far have focused on observing and interpreting DQPTs under fixed conditions. A more challenging and largely unexplored direction concerns the active control and engineering of DQPTs—specifically, how one can deliberately induce, suppress, or tailor their occurrence and characteristic features.

Building on this motivation, we investigate how squeezing can actively engineer DQPTs. A key challenge lies in defining a suitable squeezing operator: a naive construction in terms of spin operators yields only a trivial global phase and provides no genuine state control. We therefore introduce a double-mode squeezing operator directly in the diagonal Bogoliubov basis, $(k,-k)$, of the pre-quench Hamiltonian. Its spin representation—obtained via inverse Jordan-Wigner transformation—takes the form of a nonlocal string operator, confirming its nontrivial action. Armed with this tool, the dynamics separate into two regimes by the operator’s particle-hole symmetry (PHS). When PHS is broken, DQPTs become broadly tunable. Remarkably, when PHS is preserved at a special strength $r=\pi/4$, DQPTs enter a universal regime independent of the quench path. To understand this, we analyze the associated dynamical topological order parameter (DTOP) and find a purely geometric evolution with sharp $\pi$-jumps, as the dynamical phase vanishes. Further examination of the entanglement entropy in each $(k,-k)$ pair reveals that at $r=\pi/4$ (with $\phi=0$), the squeezed state remains maximally entangled at all times. Moreover, the critical momenta governing DQPTs coincide exactly with the modes that saturate maximal entanglement in the general case. Thus, our work establishes double-mode squeezing in the Bogoliubov basis as a potent and feasible protocol for controlling DQPTs.

\textit{Models.}---
We consider a one-dimensional spin-$\frac{1}{2}$ XY chain in a transverse field, a paradigmatic model for studying quantum criticality and nonequilibrium dynamics. Its Hamiltonian reads
\begin{equation}
H = -\frac{1}{2}\sum_{n=1}^{N}\left( \frac{1+\gamma}{2}\sigma_{n}^{x}\sigma_{n+1}^{x} + \frac{1-\gamma}{2}\sigma_{n}^{y}\sigma_{n+1}^{y} + h\sigma_{n}^{z} \right),
\end{equation}
where $\sigma_{n}^{\alpha}$ ($\alpha=x,y,z$) are Pauli matrices at site $n$, $\gamma$ is the anisotropy parameter, and $h$ is the transverse field strength. The model encompasses the transverse-field Ising chain ($\gamma=1$) and the XX model ($\gamma=0$). Imposing periodic boundary conditions and applying a Fourier transformation, the Hamiltonian can be recast in momentum space as $H = \sum_{k>0} \Psi_{k}^{\dagger} \mathcal{H}_{k} \Psi_{k}$, with the Nambu spinor $\Psi_{k} = (c_{k}, c_{-k}^{\dagger})^{T}$ and the $2\times2$ Bloch Hamiltonian
\begin{equation}
\mathcal{H}_{k} = - (h+\cos k)~\sigma^{z} - \gamma\sin k~\sigma^{y}.
\end{equation}
This form manifestly respects PHS: $\mathcal{C} \mathcal{H}_{k} \mathcal{C}^{-1} = -\mathcal{H}_{-k}$, where the PHS operator acts as $\mathcal{C} = \sigma^{x} \mathcal{K}$ ($\mathcal{K}$ denotes complex conjugation).

Diagonalization is achieved via a Bogoliubov transformation, $c_{k}=\cos{\theta_{k}}\eta_{k}+i\sin{\theta_{k}}\eta_{-k}^{\dag}$
with the Bogoliubov angle $\theta_k$ determined by $\tan 2\theta_{k} = \gamma\sin k / (h+\cos k)$. In terms of the Bogoliubov quasiparticles $\eta_{k}$, the Hamiltonian takes the diagonal form
\begin{equation}
H = \sum_{k} \varepsilon_{k} \left( \eta_{k}^{\dagger}\eta_{k} - \tfrac{1}{2} \right),
\end{equation}
where the excitation spectrum is $\varepsilon_{k} = \sqrt{(h+\cos k)^{2} + \gamma^{2}\sin^{2}k}$.

\textit{Double-modes squeezing operator.}---
As discussed in the introduction, a naive definition of squeezing directly in spin-operator space yields only trivial global phases, precluding any genuine control over the quantum state. A physically effective squeezing operator must instead act on the system's fundamental excitations. For the XY chain, these are the Bogoliubov quasiparticles. Motivated by the inherent coupling between Cooper-pair modes $k$ and $-k$, we therefore define the double-modes squeezing operator in the quasiparticle ($\eta_k$, $\eta_k^\dagger$) space \cite{Ma2011PhysRep}:
\begin{equation}
  \hat{S}(\xi) = \prod_{k>0}\hat{S}_{k}(\xi) = \prod_{k>0}\exp{(\xi^{*}\eta_{-k}\eta_{k}-\xi\eta_{k}^{\dag}\eta_{-k}^{\dag})},
\end{equation}
where the complex parameter $\xi = r e^{i\phi}$ encodes the squeezing strength $r\ge 0$ and the direction $\phi$. Transforming the squeezing operator back to the spin representation via the inverse Jordan–Wigner transformation yields a highly nonlocal structure:
\begin{equation}
  \hat{S} = \exp\Bigg[\sum_{x<y}\Big(J_{xy}\sigma_x^+\sigma_y^+\prod_{m=x}^{y-1}(-\sigma_m^z)-\mathrm{H.c.}\Big)\Bigg],
\end{equation}
where the pairing amplitude
\begin{equation}
  J_{xy}= \frac{1}{2\pi}\int_0^{\pi}\theta_{k}\sin\big[k(y-x)\big]dk
\end{equation}
encodes the range and strength of the squeezing process. This form makes explicit that double-mode squeezing in the spin language is a coherent superposition of long-range spin-pair creation mediated by a string of $\sigma^z$ operators, which faithfully encodes the underlying fermionic statistics. Thus, the spin representation reveals how squeezing—originally defined in the Bogoliubov quasiparticle basis—acts as a nonlocal entangling operation in real space. 

To analyze its action, we restrict attention to the two-dimensional subspace spanned by ${|0_k0_{-k}\rangle, \eta_{k}^{\dag}\eta_{-k}^{\dag}|0_k0_{-k}\rangle \equiv |1_k1_{-k}\rangle }$. Within this subspace, the generator $\hat{M}_k \equiv \xi^{*} \eta_{-k}\eta_{k} - \xi \eta_{k}^{\dag}\eta_{-k}^{\dag}$ takes the matrix form
\begin{equation}
M_k =
\begin{pmatrix}
0 & \xi^* \\
-\xi & 0
\end{pmatrix}
=
\begin{pmatrix}
0 & r e^{-i\phi} \\
-re^{i\phi} & 0
\end{pmatrix}.
\end{equation}
Noting that $M^{2} = -r^{2}I$, the exponential can be evaluated via the Taylor series, yielding
\begin{equation}
\hat{S}_k(\xi) = e^{M_k} = \cos rI + \frac{\sin r}{r}M_k
= \begin{pmatrix}
    \cos{r} & e^{i\phi}\sin{r} \\
      -e^{-i\phi}\sin{r} & \cos{r}
\end{pmatrix}.
\end{equation}
Applying $S_k(\xi)$ to the vacuum state gives the squeezed state
\begin{equation}
  \hat{S}_{k}(\xi)|0_{k}0_{-k}\rangle = \cos{r}|0_{k}0_{-k}\rangle - e^{i\phi}\sin{r}|1_{k}1_{-k}\rangle.
\end{equation}
This result reveals a key periodicity in the squeezing strength $r$: the state satisfies $|\psi_k^s(r+2\pi)\rangle = |\psi_k^s(r)\rangle$. The periodicity originates from the fact that, due to the Pauli exclusion principle, the system can only populate the two states $|0_k0_{-k}\rangle$ and $|1_k1_{-k}\rangle$. Consequently, the squeezing dynamics within each $(k,-k)$ subspace are described by a $\mathrm{SU}(2)$ group, which naturally exhibits a $2\pi$ periodicity in its parameter space.

Finally, we examine how the squeezing operator behaves under PHS. Using $\mathcal{C}\eta_k\mathcal{C}^{-1} = \eta_{-k}^\dagger$ and $\mathcal{C} i \mathcal{C}^{-1} = -i$ (with $\mathcal{C}=\sigma^x\mathcal{K}$ in the Nambu representation), we obtain
\begin{equation}
\mathcal{C} S_k(\xi) \mathcal{C}^{-1} = S_k(\xi^{*}).
\end{equation}
Hence, $S_k(\xi)$ preserves PHS only if $\xi = \xi^{*}$, i.e., when $\phi = 0$ or $\pi$ (real squeezing parameter).

\begin{figure}
    \centering
    \includegraphics[width=1\linewidth]{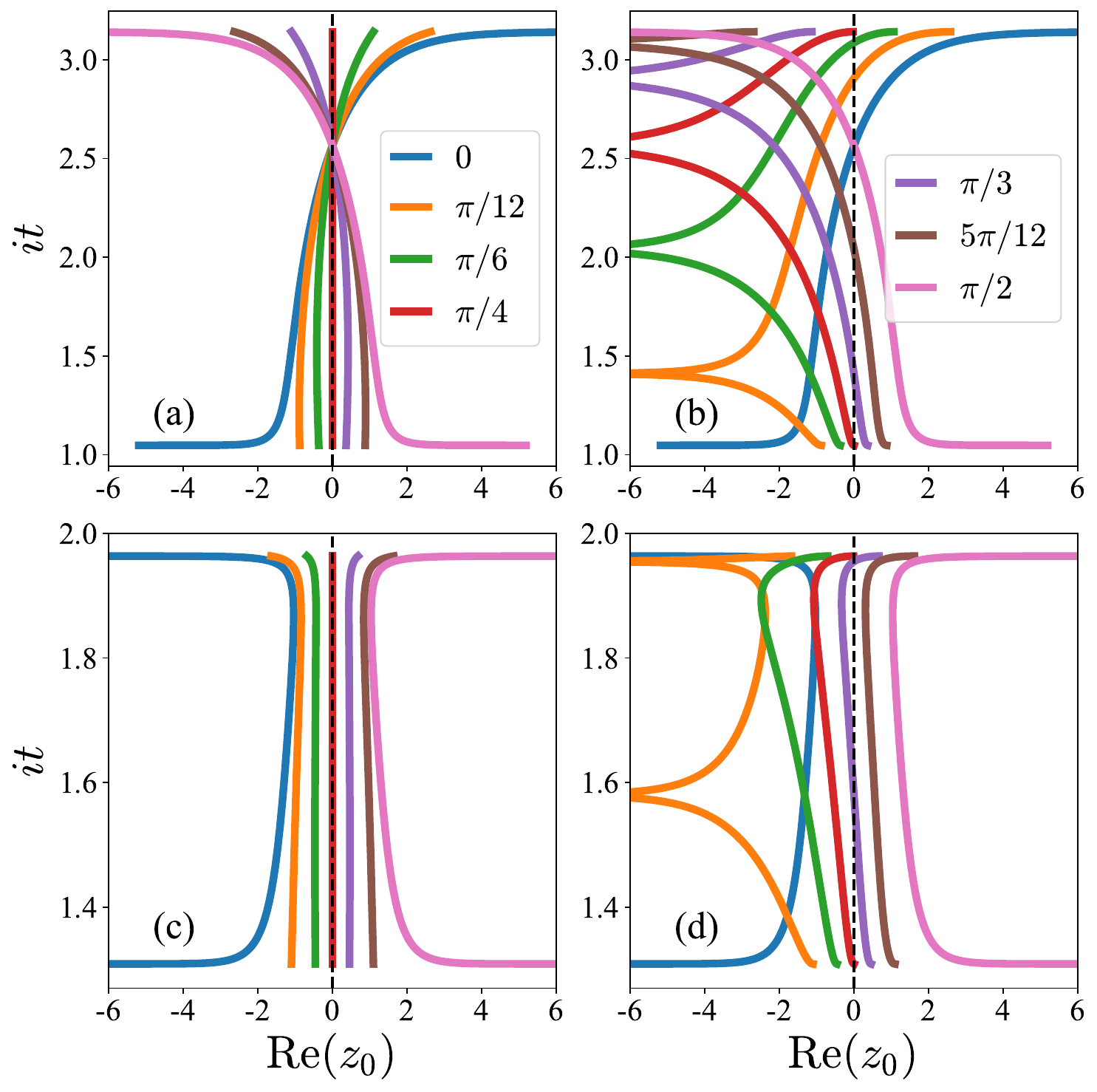}
    \caption{Fisher zeros $z_{0}$ of the Loschmidt amplitude in the complex time plane for varying squeezing strengths $r, r \in [0, \frac{\pi}{2}]$. In panels (a) and (c), the squeezing direction is set to $\phi = 0$, corresponding to the presence of PHS. In contrast, panels (b) and (d) depict PHS-broken cases with $\phi = \frac{\pi}{3}$. For (a) and (b), the quench paths cross the QPT, from $h_{0} = 1.5$ to $h_{1} = 0.5$ with $\gamma = 1$, while (c) and (d) represent quenches in the FM$_{x}$ phase, from $h_{0} = 0.8$ to $h_{1} = 0.2$ (also with $\gamma = 1$). All panels use identical line labels for consistency.}
    \label{fig: fisher.zero}
\end{figure}

\textit{Dynamical quantum phase transitions.}---We now analyze the nonequilibrium protocol via a quantum quench. The system is initialized in the ground state of the Hamiltonian $H_{0} = H(h_{0}, \gamma_{0})$, and subsequently squeezed using the operator $S(\xi)$, denoted as $|\psi_{0}\rangle = |\psi^{s}\rangle$. At $t = 0$, the system parameters are abruptly changed to $(h_{1}, \gamma_{1})$, corresponding to the post-quench Hamiltonian $\tilde{H} = H(h_{1}, \gamma_{1})$. The Loschmidt amplitude under squeezing is then given by 
\begin{equation}
    \mathcal{G}^{s}(t) = \langle\psi^{s}|\psi(t)\rangle = \langle\psi^{s}|e^{-i\tilde{H}t}|\psi^{s}\rangle.
\end{equation}
To compute this amplitude, we express the squeezed state in terms of eigenstates of $\tilde{H}$:
\begin{equation}
  |\psi_{k}^{s}\rangle = A_{k}|\tilde{0}_{k}\tilde{0}_{-k}\rangle + B_{k}|\tilde{1}_{k}\tilde{1}_{-k}\rangle,
\end{equation}
with 
\begin{eqnarray}
  A_{k} &=& \cos{r}\cos{\alpha_{k}} + ie^{i\phi}\sin{r}\sin{\alpha_{k}}, \label{eq: Ak} \\
  B_{k} &=& -i\cos{r}\sin{\alpha_{k}} - e^{i\phi}\sin{r}\cos{\alpha_{k}}, \label{eq: Bk}
\end{eqnarray}
where $\alpha_{k} = \tilde{\theta}_{k}-\theta_{k}$.
The Loschmidt amplitude becomes 
\begin{equation}
  \mathcal{G}^{s}(t) = \prod_{k>0}\mathcal{G}_{k}^{s}(t) = \prod_{k>0}(|A_{k}|^{2}e^{i\tilde{\varepsilon}_{k}t} + |B_{k}|^{2}e^{-i\tilde{\varepsilon}_{k}t}).
\end{equation}
The dynamical free energy, defined via the rate function of $\mathcal{G}^{s}(t)$, serves as a diagnostic for DQPTs:
\begin{equation}
    \lambda(t) = \lim_{N\rightarrow\infty}\frac{2}{N}\sum_{k}\ln|\mathcal{G}_{k}^{s}(t)|.
\end{equation}
Nonanalytic peaks in $\lambda(t)$ correspond to Fisher zeros on the real-time axis. These zeros, which signal nonanalytic behavior of $\mathcal{G}^{s}(t)$, are given by 
\begin{equation}\label{eq: fisher.zeros}
  z_{n} = \frac{1}{2\tilde{\varepsilon}_{k}}[\ln{\frac{|B_{k}|^{2}}{|A_{k}|^{2}}} + i(2n+1)\pi],
\end{equation}
where $z = \tau + it$ represents extended complex time. The intersection of a Fisher-zero line $z_{n}$ with the real-time axis determines the critical times for DQPTs, occurring when $|A_{k}|^{2} = |B_{k}|^{2}$. From expressions for $A_{k}$ and $B_{k}$, this condition simplifies to the existence of critical wave vectors $k'$ satisfying
\begin{equation}\label{eq: condition}
    \Delta = \cos{2r}\cos{2\alpha_{k'}} - \sin{2r}\sin{2\alpha_{k'}}\sin{\phi} = 0.
\end{equation}
For $r = 0$, this reduces to the standard DQPT criterion in the XY chain \cite{Vajna2014prb, Cao2020prb}. Notably, the squeezing strength $r$ and direction $\phi$ couple nontrivially with the quench parameters. The Eq.~(\ref{eq: condition}) can be rewritten as $\sin{\phi}=\cot{2\alpha_{k'}}\cot{2r}$, which exhibits the following symmetries: (i) Discrete translation in $r:~r\rightarrow r+\frac{\pi}{2}$; (ii) Reflection in $\phi:~\phi\rightarrow \pi-\phi$; (iii) Combined reflection: $(r, \phi)\rightarrow (\frac{\pi}{2}-r, -\phi)$. 

Intriguingly, when $r = \frac{\pi}{4}$ and $\phi = 0$, Eq.~(\ref{eq: condition}) holds for all $k$, implying that the Loschmidt amplitude (actually here we obtain $\mathcal{G}^{s}(t)=\prod_{k>0}\cos{\tilde{\varepsilon}_{k}t}$) exhibits Fisher zeros exclusively on the real-time axis . Numerical results presented below corroborate these findings.

\begin{figure}
    \centering
    \includegraphics[width=1\linewidth]{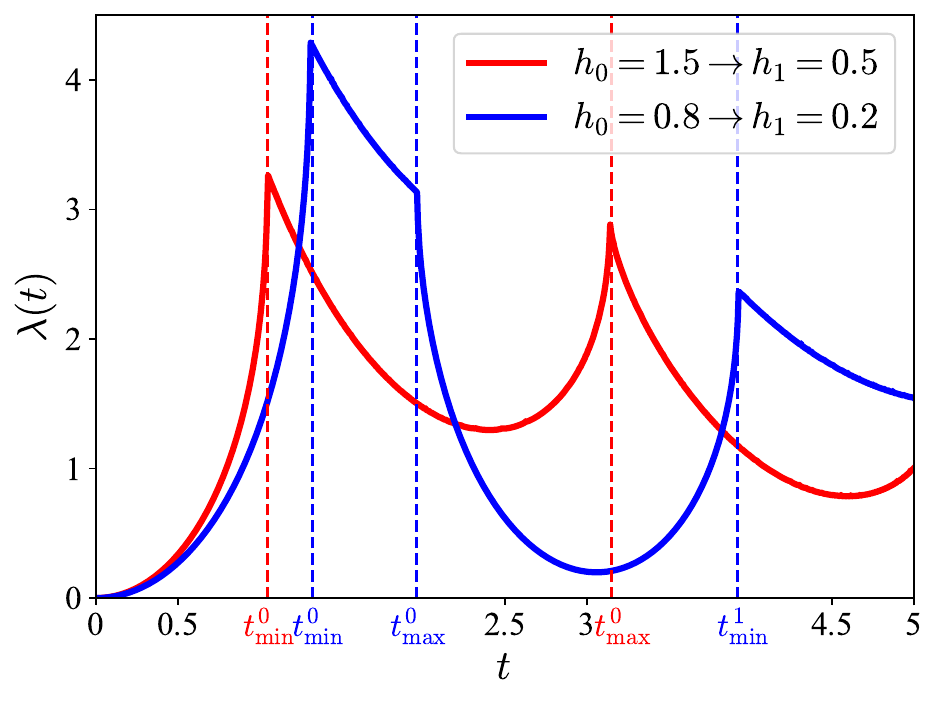}
    \caption{The rate functions for the PHS-preserved squeezing with $r = \frac{\pi}{4}$, where the critical times $t_{\mathrm{min}}^{0}$ and $t_{\mathrm{max}}^{0}$ correspond to the boundaries of Fisher zeros $z_{0}$. }
    \label{fig: rate.function}
\end{figure}

Fig.~\ref{fig: fisher.zero} shows the Fisher-zero lines $z_{0}$ in four representative scenarios, demonstrating how initial-state squeezing controls the occurrence of DQPTs. Panels (a) and (b) correspond to quenches across the quantum critical point, while (c) and (d) represent intra-phase quenches. Throughout, the anisotropy is fixed at $\gamma=1$ (the quantum Ising limit), where DQPTs conventionally emerge only for quenches crossing the critical point at $h_{c}=1$ \cite{Heyl2013PRL}.
In panel (a), where the squeezing operator preserves PHS, we find that for all $r \neq \pi/4$, the Fisher-zero lines intersect the real-time axis at precisely the same points as in the unsqueezed case ($r=0$). This shows that squeezing leaves the critical times unchanged unless $r$ reaches the special value $\pi/4$. At $r=\pi/4$, however, the Fisher zeros collapse onto the entire real-time axis, forming a continuous line segment, in full agreement with our analytical prediction. A similar universal collapse is observed for the intra-phase quench in panel (c) at $r=\pi/4$ and $\phi=0$.
Analysis of the corresponding dynamical free energy (see Fig.~\ref{fig: rate.function}) further reveals that nonanalytic peaks in the rate function appear only at the endpoints of real-time Fisher-zero segments, whereas zeros lying strictly within the interval do not produce singularities (For finite systems, multiple small peaks may appear within $[t_{\mathrm{min}}^{0}, t_{\mathrm{max}}^{0}]$, but these are smoothed out in the thermodynamic limit.) These results confirm that setting $r=\pi/4$ and $\phi=0$ universally induces DQPTs, irrespective of whether the quench crosses the critical point or remains within the same phase.

\begin{figure}
    \centering
    \includegraphics[width=1\linewidth]{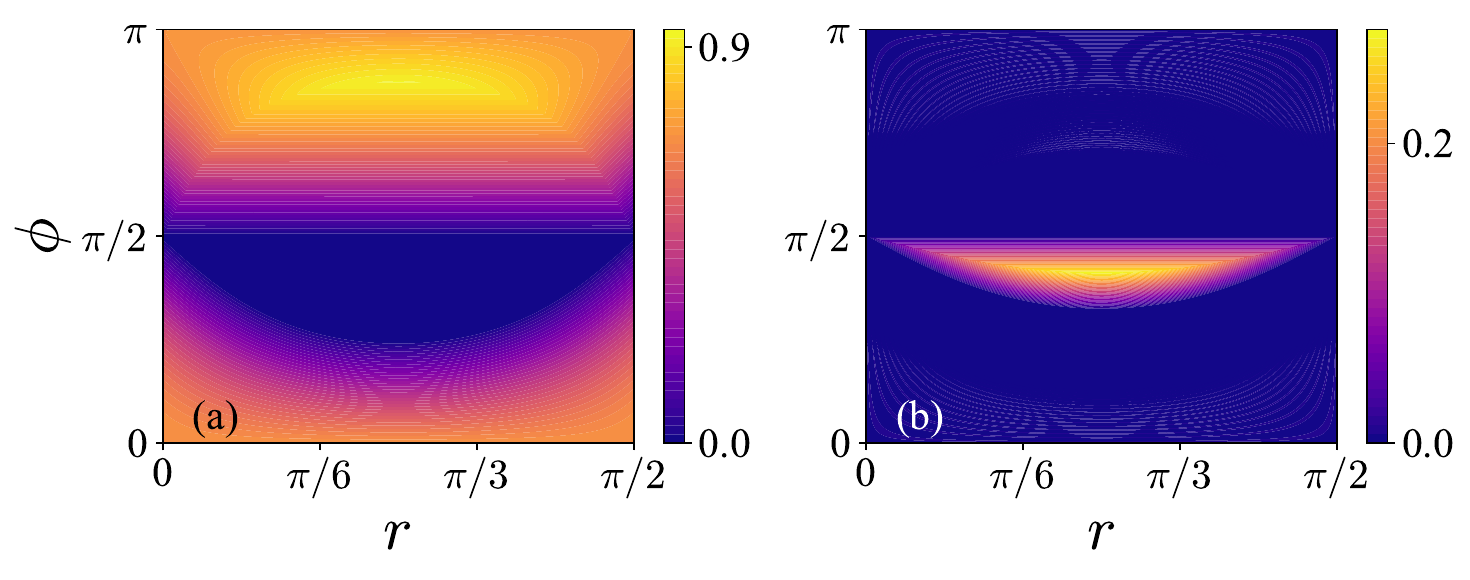}
    \caption{The contour plot of condition $\Delta(r, \phi) = \min{|\cos{2r}\cos{2\alpha_{k}}-\sin{2r}\sin{2\alpha_{k}}\sin{\phi}|}$ obtained by scanning squeezing parameters $(r,\phi)$, where the quench path for (a) is from $h_{0}=0.8$ to $h_{1}=0.2$ with $\gamma=1$ (Ising limit), and for (b) from $h_{0}=0.2$ to $h_{1}=0.8$ with $\gamma=0.1$ (XX limit).}
    \label{fig: Delta.criterion}
\end{figure}

In contrast, for the PHS-broken case [see Fig.~\ref{fig: fisher.zero}~(b)], the intersections between the Fisher-zero lines and real-time axis shift continuously with the squeezing strength, indicating that the critical times of DQPTs can be tuned by varying $r$. By tuning $r$, we can obtain any desired critical time $t_{c}^{0}$ in time interval $[t_{\mathrm{min}}^{0}, t_{\mathrm{max}}^{0}]$. Remarkably, for intra-phase quenches---where no DQPT occurs in the absence of squeezing---Fig.~\ref{fig: fisher.zero}~(d) reveals that certain values of $r$ cause the Fisher-zero lines to cross the real-time axis. Our results demonstrate that, in addition to the special case of $r=\frac{\pi}{4}, \phi=0$ with PHS, DQPTs can also be induced in quenches within the same phase by appropriately choosing the squeezing parameters under PHS-broken condition. This highlights the power of double-mode squeezing as a versatile tool for generating and tailoring dynamical criticality.

To quantitatively determine the range of squeezing parameters $(r, \phi)$ that induce DQPTs in the PHS-broken regime, we introduce a motivated criterion $\Delta$ based on Eq.~(\ref{eq: condition}):
\begin{equation}
    \Delta(r,\phi) = \min{|\cos{2r}\cos{2\alpha_{k}}-\sin{2r}\sin{2\alpha_{k}}\sin{\phi}|}.
\end{equation}
By scanning over $r$ and $\phi$, we identify parameter regions with $\Delta=0$, which correspond to the presence of DQPTs. Fig.~\ref{fig: Delta.criterion}~(a) shows the contourf plot of $\Delta(r, \phi)$ for the intra-phase quench in the quantum Ising limit ($\gamma=1$). A contiguous region where $\Delta = 0$ is clearly observed, confirming that DQPTs can be reliably induced by tuning the squeezing parameters within this regime. Furthermore, in the XY chain near the XX limit $\gamma \rightarrow 0$, intra-phase quenches are known to exhibit DQPTs even without squeezing \cite{Vajna2014prb}. We thus scan $\Delta(r, \phi)$ for such a quench protocol, with results displayed in Fig.~\ref{fig: Delta.criterion}~(b). A central region in parameter space shows $\Delta > 0$, indicating that initial-state squeezing with parameters in this region suppresses DQPTs that would otherwise occur. Together, these phase diagrams demonstrate the bidirectional control enabled by PHS-broken squeezing: it can either induce DQPTs where none were present, or suppress them where they would normally appear.

\begin{figure}
    \centering
    \includegraphics[width=1\linewidth]{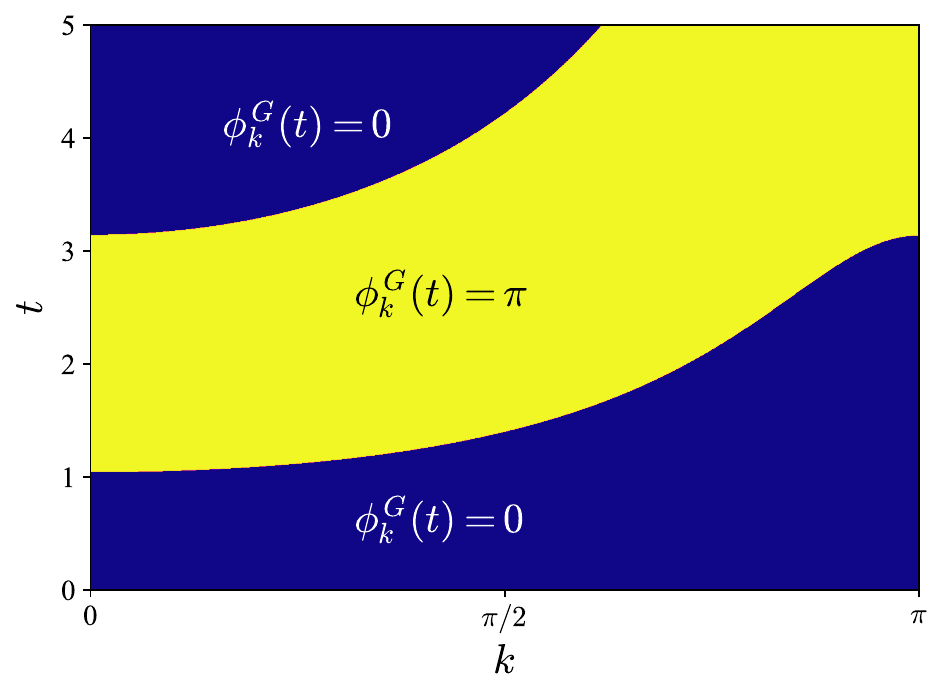}
    \caption{Pancharatnam geometric phase $\phi_{k}^{G}(t)$ in the system under PHS-preserved squeezing with $r = \frac{\pi}{4}$. }
    \label{fig: geometric.phase}
\end{figure}

\textit{Dynamical topological phase transitions.---}
While DQPTs lack a local order parameter—in contrast to equilibrium symmetry-breaking phase transitions—they can be characterized by a DTOP \cite{Budich2016PRB}, derived from the Pancharatnam geometric phase of the Loschmidt amplitude. The DTOP is integer-quantized and changes abruptly at critical times, thereby tracking the topological structure of the time-evolving quantum state.

To analyze this in our squeezed-quench setting, we express the Loschmidt amplitude in polar form, $\mathcal{G}_{k}^{s}(t) = r_{k}e^{i\phi_{k}(t)}$. The total phase $\phi_{k}(t)$ decomposes into a geometric (Pancharatnam) part $\phi_{k}^{G}(t)$ and a dynamical part $\phi_{k}^{\text{dyn}}(t)$ arising from energy accumulation. In the presence of squeezing, the dynamical phase is given by
\begin{equation}
\phi_{k}^{\text{dyn}}(t) = \big(|A_{k}|^{2} - |B_{k}|^{2}\big)\tilde{\varepsilon}_{k}t .
\end{equation}
Focusing on the special squeezing parameters $r=\pi/4$ and $\phi=0$, condition (\ref{eq: condition}) implies $|A_{k}|^{2} - |B_{k}|^{2} = 0$ for all $k$. Consequently, the dynamical phase vanishes identically, and the phase evolution becomes purely geometric: $\phi_{k}^{G}(t) = \phi_{k}(t)$. In this regime, the Loschmidt amplitude simplifies to $\mathcal{G}_{k}^{s}(t) = \cos(\tilde{\varepsilon}_{k}t)$, and the geometric phase exhibits sharp $\pi$-jumps:
\begin{equation}
\phi_{k}^{G}(t) =
\begin{cases}
0, & \cos(\tilde{\varepsilon}_{k}t) > 0, \\
\pi, & \cos(\tilde{\varepsilon}_{k}t) < 0,
\end{cases}
\end{equation}
directly signaling a nontrivial DTOP. This result reveals a distinct physical picture induced by squeezing: at $r=\pi/4$ and $\phi=0$, the many-body dynamics are stripped of all dynamical-phase contributions, leaving a purely geometric evolution that governs the DQPT. Squeezing thus acts as a tuning knob that not only triggers universal DQPTs, but also routes the system into a regime where topological signatures emerge from geometry alone—a scenario qualitatively different from conventional quenches where dynamical and geometric phases intertwine.

\textit{Double-mode entropy induced by squeezing.---}
The special squeezing parameters $r=\pi/4$ and $\phi=0$ yield more than universal DQPTs; they correspond to a maximally entangled initial state in each $(k,-k)$ subspace. To quantify this, we examine the reduced von Neumann entropy. The pure squeezed state $|\psi_k^s\rangle$ in the double-mode subspace $(k,-k)$ is described by the density matrix $\rho_{k,-k}=|\psi_k^s\rangle\langle\psi_k^s|$. Tracing out the $-k$ mode gives the reduced density matrix for mode $k$,
\begin{equation}
\rho_k = \operatorname{Tr}_{-k} \rho_{k,-k} = \operatorname{diag}\big(|A_k|^2, |B_k|^2\big),
\end{equation}
with $|A_k|^2+|B_k|^2=1$. Setting $|A_k|^2 = (1+\Delta)/2$ and $|B_k|^2 = (1-\Delta)/2$ identifies $\Delta$ as the quantity appearing in condition (\ref{eq: condition}). The von Neumann entropy then reads
\begin{equation}
\mathcal{S}_k = -\operatorname{Tr}(\rho_k\ln\rho_k)
= -\frac{1+\Delta}{2}\ln\frac{1+\Delta}{2} - \frac{1-\Delta}{2}\ln\frac{1-\Delta}{2}.
\end{equation}
Crucially, $\mathcal{S}_{k}$ reaches its maximum value $\mathcal{S}_{k,\max}= \ln 2$ if and only if $\Delta=0$, which coincides precisely with condition (\ref{eq: condition}). This equivalence reveals two fundamental insights:
\begin{enumerate}
\item The parameters $r=\pi/4$, $\phi=0$ produce a maximally entangled squeezed state ($\mathcal{S}_k=\ln 2$) in each $(k,-k)$ pair.
\item The critical wave vectors that govern the DQPTs after squeezing are exactly those modes that attain maximal entanglement ($\Delta=0$).
\end{enumerate}
Thus, at $r=\pi/4$, $\phi=0$, the universal emergence of DQPTs is intrinsically linked to the formation of maximum entanglement between momentum‑paired modes. The Fisher‑zero collapse and the subsequent nonanalyticities in the rate function are direct dynamical signatures of this underlying entanglement saturation.

\textit{Conclusions.}---
In summary, we have established initial-state double-mode squeezing as a powerful protocol for controlling DQPTs. Using the transverse-field XY chain, we demonstrate two complementary control mechanisms. When squeezing breaks PHS, DQPTs become fully tunable—critical times can be shifted arbitrarily, and transitions can be induced or suppressed at will, overriding the conventional critical-point criterion.
More profoundly, we find that squeezing which preserves particle-hole symmetry uncovers a fundamental connection between entanglement and dynamical criticality. At the special squeezing strength $r=\pi/4$, all Fisher zeros collapse onto the real-time axis, leading to universal DQPTs independent of the quench path. Crucially, this universal regime corresponds exactly to maximally entangled $(k,-k)$ mode pairs, with the critical momenta of the DQPTs matching those modes where the double-mode von Neumann entropy saturates at $\ln{2}$. Here, the dynamical phase vanishes identically, leaving a purely geometric evolution characterized by sharp $\pi$-jumps in the Pancharatnam phase and a distinct dynamical topological order parameter.
Thus, our work reveals that the universal nonanalytic signatures in quench dynamics are direct manifestations of entanglement saturation in the initial state. By bridging quantum information (double-mode entanglement) with nonequilibrium critical phenomena (DQPTs), we provide both a versatile control protocol and a deeper conceptual framework for engineering dynamical quantum phases in synthetic quantum simulators.

\begin{acknowledgments}
  K.C. was funded by Basic Research Program of Jiangsu (Grant No.~BK20250886).  J.W. was supported by the National Natural Science Foundation of China (Grant No.~11875047). S.C. was supported by National Key Research and Development Program of China (Grant No.~2021YFA1402104) and the National Natural Science Foundation under Grants No.~12474287 and No.~T2121001.
\end{acknowledgments}

\bibliography{reference}

@PREAMBLE{
 "\providecommand{\noopsort}[1]{}" 
 # "\providecommand{\singleletter}[1]{#1}%" 
}

@article{Taylor2016PhysRep,
title = {Quantum metrology and its application in biology},
journal = {Physics Reports},
volume = {615},
pages = {1-59},
year = {2016},
note = {Quantum metrology and its application in biology},
issn = {0370-1573},
doi = {https://doi.org/10.1016/j.physrep.2015.12.002},
url = {https://www.sciencedirect.com/science/article/pii/S0370157315005001},
author = {Michael A. Taylor and Warwick P. Bowen}
}

@article{Taylor2013NatPhoto,
   author = {Taylor, Michael A. and Janousek, Jiri and Daria, Vincent and Knittel, Joachim and Hage, Boris and Bachor, Hans- A. and Bowen, Warwick P.},
   title = {Biological measurement beyond the quantum limit},
   journal = {Nature Photonics},
   volume = {7},
   number = {3},
   pages = {229-233},
   ISSN = {1749-4893},
   DOI = {10.1038/nphoton.2012.346},
   url = {https://doi.org/10.1038/nphoton.2012.346},
   year = {2013},
   type = {Journal Article}
}

@article{Qvarfort2018nc,
   author = {Qvarfort, Sofia and Serafini, Alessio and Barker, P. F. and Bose, Sougato},
   title = {Gravimetry through non-linear optomechanics},
   journal = {Nature Communications},
   volume = {9},
   number = {1},
   pages = {3690},
   ISSN = {2041-1723},
   DOI = {10.1038/s41467-018-06037-z},
   url = {https://doi.org/10.1038/s41467-018-06037-z},
   year = {2018},
   type = {Journal Article}
}

@article{Simon2022SciAdv,
author = {Simon Templier  and Pierrick Cheiney  and Quentin d’Armagnac de Castanet  and Baptiste Gouraud  and Henri Porte  and Fabien Napolitano  and Philippe Bouyer  and Baptiste Battelier  and Brynle Barrett },
title = {Tracking the vector acceleration with a hybrid quantum accelerometer triad},
journal = {Science Advances},
volume = {8},
number = {45},
pages = {eadd3854},
year = {2022},
doi = {10.1126/sciadv.add3854},
URL = {https://www.science.org/doi/abs/10.1126/sciadv.add3854}
}

@article{Morris2015nc,
   author = {Morris, Peter A. and Aspden, Reuben S. and Bell, Jessica E. C. and Boyd, Robert W. and Padgett, Miles J.},
   title = {Imaging with a small number of photons},
   journal = {Nature Communications},
   volume = {6},
   number = {1},
   pages = {5913},
   ISSN = {2041-1723},
   DOI = {10.1038/ncomms6913},
   url = {https://doi.org/10.1038/ncomms6913},
   year = {2015},
   type = {Journal Article}
}

@article{Anders1999PRL,
  title = {Spin-Spin Interaction and Spin Squeezing in an Optical Lattice},
  author = {S\o{}rensen, Anders and M\o{}lmer, Klaus},
  journal = {Phys. Rev. Lett.},
  volume = {83},
  issue = {11},
  pages = {2274--2277},
  numpages = {0},
  year = {1999},
  month = {Sep},
  publisher = {American Physical Society},
  doi = {10.1103/PhysRevLett.83.2274},
  url = {https://link.aps.org/doi/10.1103/PhysRevLett.83.2274}
}

@article{Andre2004PRL,
  title = {Stability of Atomic Clocks Based on Entangled Atoms},
  author = {Andr\'e, A. and S\o{}rensen, A. S. and Lukin, M. D.},
  journal = {Phys. Rev. Lett.},
  volume = {92},
  issue = {23},
  pages = {230801},
  numpages = {4},
  year = {2004},
  month = {Jun},
  publisher = {American Physical Society},
  doi = {10.1103/PhysRevLett.92.230801},
  url = {https://link.aps.org/doi/10.1103/PhysRevLett.92.230801}
}

@article{Liu2025PRL,
  title = {Enhancing Dynamic Range of Sub-Standard-Quantum-Limit Measurements via Quantum Deamplification},
  author = {Liu, Qi and Xue, Ming and Radzihovsky, Matthew and Li, Xinwei and Vasilyev, Denis V. and Wu, Ling-Na and Vuleti\ifmmode \acute{c}\else \'{c}\fi{}, Vladan},
  journal = {Phys. Rev. Lett.},
  volume = {135},
  issue = {4},
  pages = {040801},
  numpages = {8},
  year = {2025},
  month = {Jul},
  publisher = {American Physical Society},
  doi = {10.1103/25ds-9724},
  url = {https://link.aps.org/doi/10.1103/25ds-9724}
}

@article{Zou2025PRL,
  title = {Enhancing Quantum Metrology by Quantum Resonance Dynamics},
  author = {Zou, Zhixing and Gong, Jiangbin and Chen, Weitao},
  journal = {Phys. Rev. Lett.},
  volume = {134},
  issue = {23},
  pages = {230802},
  numpages = {7},
  year = {2025},
  month = {Jun},
  publisher = {American Physical Society},
  doi = {10.1103/lkrt-lvng},
  url = {https://link.aps.org/doi/10.1103/lkrt-lvng}
}

@article{Bhattacharyya2024PRA,
  title = {Enhancing precision of atomic clocks by tuning disorder in accessories},
  author = {Bhattacharyya, Aparajita and Ghoshal, Ahana and Sen, Ujjwal},
  journal = {Phys. Rev. A},
  volume = {110},
  issue = {1},
  pages = {012620},
  numpages = {16},
  year = {2024},
  month = {Jul},
  publisher = {American Physical Society},
  doi = {10.1103/PhysRevA.110.012620},
  url = {https://link.aps.org/doi/10.1103/PhysRevA.110.012620}
}

@article{Finkelstein2024nature,
   author = {Finkelstein, Ran and Tsai, Richard Bing-Shiun and Sun, Xiangkai and Scholl, Pascal and Direkci, Su and Gefen, Tuvia and Choi, Joonhee and Shaw, Adam L. and Endres, Manuel},
   title = {Universal quantum operations and ancilla-based read-out for tweezer clocks},
   journal = {Nature},
   volume = {634},
   number = {8033},
   pages = {321-327},
   ISSN = {1476-4687},
   DOI = {10.1038/s41586-024-08005-8},
   url = {https://doi.org/10.1038/s41586-024-08005-8},
   year = {2024},
   type = {Journal Article}
}

@article{Marciniak2022nature,
   author = {Marciniak, Christian D. and Feldker, Thomas and Pogorelov, Ivan and Kaubruegger, Raphael and Vasilyev, Denis V. and van Bijnen, Rick and Schindler, Philipp and Zoller, Peter and Blatt, Rainer and Monz, Thomas},
   title = {Optimal metrology with programmable quantum sensors},
   journal = {Nature},
   volume = {603},
   number = {7902},
   pages = {604-609},
   ISSN = {1476-4687},
   DOI = {10.1038/s41586-022-04435-4},
   url = {https://doi.org/10.1038/s41586-022-04435-4},
   year = {2022},
   type = {Journal Article}
}

@article{Pezze2020PRL,
  title = {Heisenberg-Limited Noisy Atomic Clock Using a Hybrid Coherent and Squeezed State Protocol},
  author = {Pezz\`e, Luca and Smerzi, Augusto},
  journal = {Phys. Rev. Lett.},
  volume = {125},
  issue = {21},
  pages = {210503},
  numpages = {7},
  year = {2020},
  month = {Nov},
  publisher = {American Physical Society},
  doi = {10.1103/PhysRevLett.125.210503},
  url = {https://link.aps.org/doi/10.1103/PhysRevLett.125.210503}
}

@article{Agarwal1996PRA,
  title = {Ramsey spectroscopy with nonclassical light sources},
  author = {Agarwal, Girish S. and Scully, Marlan O.},
  journal = {Phys. Rev. A},
  volume = {53},
  issue = {1},
  pages = {467--470},
  numpages = {0},
  year = {1996},
  month = {Jan},
  publisher = {American Physical Society},
  doi = {10.1103/PhysRevA.53.467},
  url = {https://link.aps.org/doi/10.1103/PhysRevA.53.467}
}

@article{Xu1999PRA,
  title = {State-selective Rabi and Ramsey magnetic resonance line shapes},
  author = {Xu, G. and Heinzen, D. J.},
  journal = {Phys. Rev. A},
  volume = {59},
  issue = {2},
  pages = {R922--R925},
  numpages = {0},
  year = {1999},
  month = {Feb},
  publisher = {American Physical Society},
  doi = {10.1103/PhysRevA.59.R922},
  url = {https://link.aps.org/doi/10.1103/PhysRevA.59.R922}
}

@article{Sanchez2021PRL,
  title = {Squeezed Lasing},
  author = {S\'anchez Mu\~noz, Carlos and Jaksch, Dieter},
  journal = {Phys. Rev. Lett.},
  volume = {127},
  issue = {18},
  pages = {183603},
  numpages = {7},
  year = {2021},
  month = {Oct},
  publisher = {American Physical Society},
  doi = {10.1103/PhysRevLett.127.183603},
  url = {https://link.aps.org/doi/10.1103/PhysRevLett.127.183603}
}

@article{Kenan2013OptLett,
author = {Kenan Qu and G. S. Agarwal},
journal = {Opt. Lett.},
number = {14},
pages = {2563--2565},
publisher = {Optica Publishing Group},
title = {Ramsey spectroscopy with squeezed light},
volume = {38},
month = {Jul},
year = {2013},
url = {https://opg.optica.org/ol/abstract.cfm?URI=ol-38-14-2563},
doi = {10.1364/OL.38.002563}
}

@article{Walls1981PhysLettA,
title = {Enhanced sensitivity of a gravitational wave detector},
journal = {Physics Letters A},
volume = {85},
number = {2},
pages = {118-120},
year = {1981},
issn = {0375-9601},
doi = {https://doi.org/10.1016/0375-9601(81)90238-3},
url = {https://www.sciencedirect.com/science/article/pii/0375960181902383},
author = {D.F. Walls and P. Zoller}
}

@article{Goda2008NatPhys,
   author = {Goda, K. and Miyakawa, O. and Mikhailov, E. E. and Saraf, S. and Adhikari, R. and McKenzie, K. and Ward, R. and Vass, S. and Weinstein, A. J. and Mavalvala, N.},
   title = {A quantum-enhanced prototype gravitational-wave detector},
   journal = {Nature Physics},
   volume = {4},
   number = {6},
   pages = {472-476},
   ISSN = {1745-2481},
   DOI = {10.1038/nphys920},
   url = {https://doi.org/10.1038/nphys920},
   year = {2008},
   type = {Journal Article}
}

@article{Wang2022SciRep,
   author = {Wang, Liu and Xie, Fang and Zhang, Yong and Xiao, Min and Liu, Fang},
   title = {Adaptive optical phase estimation for real-time sensing of fast-varying signals},
   journal = {Scientific Reports},
   volume = {12},
   number = {1},
   pages = {21745},
   ISSN = {2045-2322},
   DOI = {10.1038/s41598-022-26329-1},
   url = {https://doi.org/10.1038/s41598-022-26329-1},
   year = {2022},
   type = {Journal Article}
}

@article{Ma2011PhysRep,
title = {Quantum spin squeezing},
journal = {Physics Reports},
volume = {509},
number = {2},
pages = {89-165},
year = {2011},
issn = {0370-1573},
doi = {https://doi.org/10.1016/j.physrep.2011.08.003},
url = {https://www.sciencedirect.com/science/article/pii/S0370157311002201},
author = {Jian Ma and Xiaoguang Wang and C.P. Sun and Franco Nori}
}

@article{Wang2003PRA,
  title = {Spin squeezing and pairwise entanglement for symmetric multiqubit states},
  author = {Wang, Xiaoguang and Sanders, Barry C.},
  journal = {Phys. Rev. A},
  volume = {68},
  issue = {1},
  pages = {012101},
  numpages = {6},
  year = {2003},
  month = {Jul},
  publisher = {American Physical Society},
  doi = {10.1103/PhysRevA.68.012101},
  url = {https://link.aps.org/doi/10.1103/PhysRevA.68.012101}
}

@article{Sorensen2001PRL,
  title = {Entanglement and Extreme Spin Squeezing},
  author = {S\o{}rensen, Anders S. and M\o{}lmer, Klaus},
  journal = {Phys. Rev. Lett.},
  volume = {86},
  issue = {20},
  pages = {4431--4434},
  numpages = {0},
  year = {2001},
  month = {May},
  publisher = {American Physical Society},
  doi = {10.1103/PhysRevLett.86.4431},
  url = {https://link.aps.org/doi/10.1103/PhysRevLett.86.4431}
}

@article{Cai2025PhysRep,
title = {A review of quantum correlation sharing: The recycling of quantum correlations triggered by quantum measurements},
journal = {Physics Reports},
volume = {1098},
pages = {1-53},
year = {2025},
note = {A review of quantum correlation sharing: The recycling of quantum correlations triggered by quantum measurements},
issn = {0370-1573},
doi = {https://doi.org/10.1016/j.physrep.2024.10.003},
url = {https://www.sciencedirect.com/science/article/pii/S0370157324003600},
author = {Zinuo Cai and Changliang Ren and Tianfeng Feng and Xiaoqi Zhou and Jingling Chen}
}

@article{Liu2025PRX,
  title = {Entanglement Witness for Indistinguishable Electrons Using Solid-State Spectroscopy},
  author = {Liu, Tongtong and Xu, Luogen and Liu, Jiarui and Wang, Yao},
  journal = {Phys. Rev. X},
  volume = {15},
  issue = {1},
  pages = {011056},
  numpages = {31},
  year = {2025},
  month = {Mar},
  publisher = {American Physical Society},
  doi = {10.1103/PhysRevX.15.011056},
  url = {https://link.aps.org/doi/10.1103/PhysRevX.15.011056}
}

@article{Mazza2025PRB,
  title = {Entanglement detection in quantum materials with competing orders},
  author = {Mazza, Giacomo and Budroni, Costantino},
  journal = {Phys. Rev. B},
  volume = {111},
  issue = {10},
  pages = {L100302},
  numpages = {7},
  year = {2025},
  month = {Mar},
  publisher = {American Physical Society},
  doi = {10.1103/PhysRevB.111.L100302},
  url = {https://link.aps.org/doi/10.1103/PhysRevB.111.L100302}
}

@article{Tavakoli2024RMP,
  title = {Semidefinite programming relaxations for quantum correlations},
  author = {Tavakoli, Armin and Pozas-Kerstjens, Alejandro and Brown, Peter and Ara\'ujo, Mateus},
  journal = {Rev. Mod. Phys.},
  volume = {96},
  issue = {4},
  pages = {045006},
  numpages = {68},
  year = {2024},
  month = {Dec},
  publisher = {American Physical Society},
  doi = {10.1103/RevModPhys.96.045006},
  url = {https://link.aps.org/doi/10.1103/RevModPhys.96.045006}
}

@article{Polkovnikov2011RMP,
  title = {Colloquium: Nonequilibrium dynamics of closed interacting quantum systems},
  author = {Polkovnikov, Anatoli and Sengupta, Krishnendu and Silva, Alessandro and Vengalattore, Mukund},
  journal = {Rev. Mod. Phys.},
  volume = {83},
  issue = {3},
  pages = {863--883},
  numpages = {0},
  year = {2011},
  month = {Aug},
  publisher = {American Physical Society},
  doi = {10.1103/RevModPhys.83.863},
  url = {https://link.aps.org/doi/10.1103/RevModPhys.83.863}
}

@article{Heyl2018RepPro,
doi = {10.1088/1361-6633/aaaf9a},
url = {https://doi.org/10.1088/1361-6633/aaaf9a},
year = {2018},
month = {apr},
publisher = {IOP Publishing},
volume = {81},
number = {5},
pages = {054001},
author = {Heyl, Markus},
title = {Dynamical quantum phase transitions: a review},
journal = {Reports on Progress in Physics}
}

@article{Zvyagin2016LowTem,
    author = {Zvyagin, A. A.},
    title = {Dynamical quantum phase transitions (Review Article)},
    journal = {Low Temperature Physics},
    volume = {42},
    number = {11},
    pages = {971-994},
    year = {2016},
    month = {11},
    issn = {1063-777X},
    doi = {10.1063/1.4969869},
    url = {https://doi.org/10.1063/1.4969869}
}

@article{Heyl2019EuLett,
doi = {10.1209/0295-5075/125/26001},
url = {https://doi.org/10.1209/0295-5075/125/26001},
year = {2019},
month = {feb},
publisher = {EDP Sciences, IOP Publishing and Società Italiana di Fisica},
volume = {125},
number = {2},
pages = {26001},
author = {Heyl, Markus},
title = {Dynamical quantum phase transitions: A brief survey},
journal = {Europhysics Letters}
}

@article{Heyl2013PRL,
  title = {Dynamical Quantum Phase Transitions in the Transverse-Field Ising Model},
  author = {Heyl, M. and Polkovnikov, A. and Kehrein, S.},
  journal = {Phys. Rev. Lett.},
  volume = {110},
  issue = {13},
  pages = {135704},
  numpages = {5},
  year = {2013},
  month = {Mar},
  publisher = {American Physical Society},
  doi = {10.1103/PhysRevLett.110.135704},
  url = {https://link.aps.org/doi/10.1103/PhysRevLett.110.135704}
}

@article{Rylands2019PRB,
  title = {Loschmidt amplitude and work distribution in quenches of the sine-Gordon model},
  author = {Rylands, Colin and Andrei, Natan},
  journal = {Phys. Rev. B},
  volume = {99},
  issue = {8},
  pages = {085133},
  numpages = {19},
  year = {2019},
  month = {Feb},
  publisher = {American Physical Society},
  doi = {10.1103/PhysRevB.99.085133},
  url = {https://link.aps.org/doi/10.1103/PhysRevB.99.085133}
}

@article{Campbell2016PRB,
  title = {Criticality revealed through quench dynamics in the Lipkin-Meshkov-Glick model},
  author = {Campbell, Steve},
  journal = {Phys. Rev. B},
  volume = {94},
  issue = {18},
  pages = {184403},
  numpages = {7},
  year = {2016},
  month = {Nov},
  publisher = {American Physical Society},
  doi = {10.1103/PhysRevB.94.184403},
  url = {https://link.aps.org/doi/10.1103/PhysRevB.94.184403}
}

@article{Lupo2016PRB,
  title = {Transient Loschmidt echo in quenched Ising chains},
  author = {Lupo, Carla and Schir\'o, Marco},
  journal = {Phys. Rev. B},
  volume = {94},
  issue = {1},
  pages = {014310},
  numpages = {13},
  year = {2016},
  month = {Jul},
  publisher = {American Physical Society},
  doi = {10.1103/PhysRevB.94.014310},
  url = {https://link.aps.org/doi/10.1103/PhysRevB.94.014310}
}

@article{Marino2014PRB,
  title = {Nonequilibrium dynamics of a noisy quantum Ising chain: Statistics of work and prethermalization after a sudden quench of the transverse field},
  author = {Marino, Jamir and Silva, Alessandro},
  journal = {Phys. Rev. B},
  volume = {89},
  issue = {2},
  pages = {024303},
  numpages = {16},
  year = {2014},
  month = {Jan},
  publisher = {American Physical Society},
  doi = {10.1103/PhysRevB.89.024303},
  url = {https://link.aps.org/doi/10.1103/PhysRevB.89.024303}
}

@article{Budich2016PRB,
  title = {Dynamical topological order parameters far from equilibrium},
  author = {Budich, Jan Carl and Heyl, Markus},
  journal = {Phys. Rev. B},
  volume = {93},
  issue = {8},
  pages = {085416},
  numpages = {7},
  year = {2016},
  month = {Feb},
  publisher = {American Physical Society},
  doi = {10.1103/PhysRevB.93.085416},
  url = {https://link.aps.org/doi/10.1103/PhysRevB.93.085416}
}

@article{Tang2025PRB,
  title = {Geometry effect of dynamical quantum phase transitions at finite temperatures},
  author = {Tang, Jia-Chen and Hou, Xu-Yang and Guo, Hao},
  journal = {Phys. Rev. B},
  volume = {111},
  issue = {17},
  pages = {174310},
  numpages = {13},
  year = {2025},
  month = {May},
  publisher = {American Physical Society},
  doi = {10.1103/PhysRevB.111.174310},
  url = {https://link.aps.org/doi/10.1103/PhysRevB.111.174310}
}

@article{Jing2024PRL,
  title = {Biorthogonal Dynamical Quantum Phase Transitions in Non-Hermitian Systems},
  author = {Jing, Yecheng and Dong, Jian-Jun and Zhang, Yu-Yu and Hu, Zi-Xiang},
  journal = {Phys. Rev. Lett.},
  volume = {132},
  issue = {22},
  pages = {220402},
  numpages = {7},
  year = {2024},
  month = {May},
  publisher = {American Physical Society},
  doi = {10.1103/PhysRevLett.132.220402},
  url = {https://link.aps.org/doi/10.1103/PhysRevLett.132.220402}
}

@article{Heyl2015PRL,
  title = {Scaling and Universality at Dynamical Quantum Phase Transitions},
  author = {Heyl, Markus},
  journal = {Phys. Rev. Lett.},
  volume = {115},
  issue = {14},
  pages = {140602},
  numpages = {5},
  year = {2015},
  month = {Oct},
  publisher = {American Physical Society},
  doi = {10.1103/PhysRevLett.115.140602},
  url = {https://link.aps.org/doi/10.1103/PhysRevLett.115.140602}
}

@article{Zamani2024JPCM,
doi = {10.1088/1361-648X/ad4df9},
url = {https://doi.org/10.1088/1361-648X/ad4df9},
year = {2024},
month = {jun},
publisher = {IOP Publishing},
volume = {36},
number = {35},
pages = {355401},
author = {Zamani, Sara and Naji, J and Jafari, R and Langari, A},
title = {Scaling and universality at ramped quench dynamical quantum phase transitions},
journal = {Journal of Physics: Condensed Matter}
}

@article{Karrasch2017PRB,
  title = {Dynamical quantum phase transitions in the quantum Potts chain},
  author = {Karrasch, C. and Schuricht, D.},
  journal = {Phys. Rev. B},
  volume = {95},
  issue = {7},
  pages = {075143},
  numpages = {4},
  year = {2017},
  month = {Feb},
  publisher = {American Physical Society},
  doi = {10.1103/PhysRevB.95.075143},
  url = {https://link.aps.org/doi/10.1103/PhysRevB.95.075143}
}

@article{Vogel2017naturep,
   author = {Fl\"{a}schner, N. and Vogel, D. and Tarnowski, M. and Rem, B. S. and L\"{u}hmann, D. S. and Heyl, M. and Budich, J. C. and Mathey, L. and Sengstock, K. and Weitenberg, C.},
   title = {Observation of dynamical vortices after quenches in a system with topology},
   journal = {Nature Physics},
   volume = {14},
   number = {3},
   pages = {265-268},
   ISSN = {1745-2473},
   DOI = {10.1038/s41567-017-0013-8},
   year = {2017},
   type = {Journal Article}
}

@article{Jurcevic2017prl,
  title = {Direct Observation of Dynamical Quantum Phase Transitions in an Interacting Many-Body System},
  author = {Jurcevic, P. and Shen, H. and Hauke, P. and Maier, C. and Brydges, T. and Hempel, C. and Lanyon, B. P. and Heyl, M. and Blatt, R. and Roos, C. F.},
  journal = {Phys. Rev. Lett.},
  volume = {119},
  issue = {8},
  pages = {080501},
  numpages = {5},
  year = {2017},
  month = {Aug},
  publisher = {American Physical Society},
  doi = {10.1103/PhysRevLett.119.080501},
  url = {https://link.aps.org/doi/10.1103/PhysRevLett.119.080501}
}

@article{Chen2020pra,
  title = {Experimentally detecting dynamical quantum phase transitions in a slowly quenched Ising-chain model},
  author = {Chen, Zhe and Cui, Jin-Ming and Ai, Ming-Zhong and He, Ran and Huang, Yun-Feng and Han, Yong-Jian and Li, Chuan-Feng and Guo, Guang-Can},
  journal = {Phys. Rev. A},
  volume = {102},
  issue = {4},
  pages = {042222},
  numpages = {9},
  year = {2020},
  month = {Oct},
  publisher = {American Physical Society},
  doi = {10.1103/PhysRevA.102.042222},
  url = {https://link.aps.org/doi/10.1103/PhysRevA.102.042222}
}

@article{Muniz2020nature,
   author = {Muniz, Juan A. and Barberena, Diego and Lewis-Swan, Robert J. and Young, Dylan J. and Cline, Julia R. K. and Rey, Ana Maria and Thompson, James K.},
   title = {Exploring dynamical phase transitions with cold atoms in an optical  cavity},
   journal = {Nature},
   volume = {580},
   number = {7805},
   pages = {602-607},
   ISSN = {1476-4687},
   DOI = {10.1038/s41586-020-2224-x},
   url = {https://doi.org/10.1038/s41586-020-2224-x},
   year = {2020},
   type = {Journal Article}
}

@article{Nie2020prl,
  title = {Experimental Observation of Equilibrium and Dynamical Quantum Phase Transitions via Out-of-Time-Ordered Correlators},
  author = {Nie, Xinfang and Wei, Bo-Bo and Chen, Xi and Zhang, Ze and Zhao, Xiuzhu and Qiu, Chudan and Tian, Yu and Ji, Yunlan and Xin, Tao and Lu, Dawei and Li, Jun},
  journal = {Phys. Rev. Lett.},
  volume = {124},
  issue = {25},
  pages = {250601},
  numpages = {6},
  year = {2020},
  month = {Jun},
  publisher = {American Physical Society},
  doi = {10.1103/PhysRevLett.124.250601},
  url = {https://link.aps.org/doi/10.1103/PhysRevLett.124.250601}
}

@article{Wang2019prl,
  title = {Simulating Dynamic Quantum Phase Transitions in Photonic Quantum Walks},
  author = {Wang, Kunkun and Qiu, Xingze and Xiao, Lei and Zhan, Xiang and Bian, Zhihao and Yi, Wei and Xue, Peng},
  journal = {Phys. Rev. Lett.},
  volume = {122},
  issue = {2},
  pages = {020501},
  numpages = {6},
  year = {2019},
  month = {Jan},
  publisher = {American Physical Society},
  doi = {10.1103/PhysRevLett.122.020501},
  url = {https://link.aps.org/doi/10.1103/PhysRevLett.122.020501}
}

@article{Xu20209lightsa,
   author = {Xu, X. Y. and Wang, Q. Q. and Heyl, M. and Budich, J. C. and Pan, W. W. and Chen, Z. and Jan, M. and Sun, K. and Xu, J. S. and Han, Y. J. and Li, C. F. and Guo, G. C.},
   title = {Measuring a dynamical topological order parameter in quantum walks},
   journal = {Light-Science Applications},
   volume = {9},
   number = {1},
   ISSN = {2047-7538},
   DOI = {10.1038/s41377-019-0237-8},
   url = {https://doi.org/10.1038/s41377-019-0237-8},
   year = {2020},
   type = {Journal Article}
}

@article{Tian2020prl,
  title = {Observation of Dynamical Quantum Phase Transitions with Correspondence in an Excited State Phase Diagram},
  author = {Tian, T. and Yang, H.-X. and Qiu, L.-Y. and Liang, H.-Y. and Yang, Y.-B. and Xu, Y. and Duan, L.-M.},
  journal = {Phys. Rev. Lett.},
  volume = {124},
  issue = {4},
  pages = {043001},
  numpages = {6},
  year = {2020},
  month = {Jan},
  publisher = {American Physical Society},
  doi = {10.1103/PhysRevLett.124.043001},
  url = {https://link.aps.org/doi/10.1103/PhysRevLett.124.043001}
}

@article{Vajna2014prb,
  title = {Disentangling dynamical phase transitions from equilibrium phase transitions},
  author = {Vajna, Szabolcs and D\'ora, Bal\'azs},
  journal = {Phys. Rev. B},
  volume = {89},
  issue = {16},
  pages = {161105},
  numpages = {5},
  year = {2014},
  month = {Apr},
  publisher = {American Physical Society},
  doi = {10.1103/PhysRevB.89.161105},
  url = {https://link.aps.org/doi/10.1103/PhysRevB.89.161105}
}

@article{Schmitt2015prb,
  title = {Dynamical quantum phase transitions in the Kitaev honeycomb model},
  author = {Schmitt, Markus and Kehrein, Stefan},
  journal = {Phys. Rev. B},
  volume = {92},
  issue = {7},
  pages = {075114},
  numpages = {13},
  year = {2015},
  month = {Aug},
  publisher = {American Physical Society},
  doi = {10.1103/PhysRevB.92.075114},
  url = {https://link.aps.org/doi/10.1103/PhysRevB.92.075114}
}

@article{Karrasch2013prb,
  title = {Dynamical phase transitions after quenches in nonintegrable models},
  author = {Karrasch, C. and Schuricht, D.},
  journal = {Phys. Rev. B},
  volume = {87},
  issue = {19},
  pages = {195104},
  numpages = {8},
  year = {2013},
  month = {May},
  publisher = {American Physical Society},
  doi = {10.1103/PhysRevB.87.195104},
  url = {https://link.aps.org/doi/10.1103/PhysRevB.87.195104}
}

@article{Kriel2014prb,
  title = {Dynamical quantum phase transitions in the axial next-nearest-neighbor Ising chain},
  author = {Kriel, J. N. and Karrasch, C. and Kehrein, S.},
  journal = {Phys. Rev. B},
  volume = {90},
  issue = {12},
  pages = {125106},
  numpages = {9},
  year = {2014},
  month = {Sep},
  publisher = {American Physical Society},
  doi = {10.1103/PhysRevB.90.125106},
  url = {https://link.aps.org/doi/10.1103/PhysRevB.90.125106}
}

@article{Sharma2015prb,
  title = {Quenches and dynamical phase transitions in a nonintegrable quantum Ising model},
  author = {Sharma, Shraddha and Suzuki, Sei and Dutta, Amit},
  journal = {Phys. Rev. B},
  volume = {92},
  issue = {10},
  pages = {104306},
  numpages = {7},
  year = {2015},
  month = {Sep},
  publisher = {American Physical Society},
  doi = {10.1103/PhysRevB.92.104306},
  url = {https://link.aps.org/doi/10.1103/PhysRevB.92.104306}
}

@article{Halimeh2017prb,
  title = {Dynamical phase diagram of quantum spin chains with long-range interactions},
  author = {Halimeh, Jad C. and Zauner-Stauber, Valentin},
  journal = {Phys. Rev. B},
  volume = {96},
  issue = {13},
  pages = {134427},
  numpages = {5},
  year = {2017},
  month = {Oct},
  publisher = {American Physical Society},
  doi = {10.1103/PhysRevB.96.134427},
  url = {https://link.aps.org/doi/10.1103/PhysRevB.96.134427}
}

@article{Homrighausen2017prb,
  title = {Anomalous dynamical phase in quantum spin chains with long-range interactions},
  author = {Homrighausen, Ingo and Abeling, Nils O. and Zauner-Stauber, Valentin and Halimeh, Jad C.},
  journal = {Phys. Rev. B},
  volume = {96},
  issue = {10},
  pages = {104436},
  numpages = {7},
  year = {2017},
  month = {Sep},
  publisher = {American Physical Society},
  doi = {10.1103/PhysRevB.96.104436},
  url = {https://link.aps.org/doi/10.1103/PhysRevB.96.104436}
}

@article{Obuchi2017prb,
  title = {Complex semiclassical analysis of the Loschmidt amplitude and dynamical quantum phase transitions},
  author = {Obuchi, Tomoyuki and Suzuki, Sei and Takahashi, Kazutaka},
  journal = {Phys. Rev. B},
  volume = {95},
  issue = {17},
  pages = {174305},
  numpages = {11},
  year = {2017},
  month = {May},
  publisher = {American Physical Society},
  doi = {10.1103/PhysRevB.95.174305},
  url = {https://link.aps.org/doi/10.1103/PhysRevB.95.174305}
}

@article{Halimeh2020prr,
  title = {Quasiparticle origin of dynamical quantum phase transitions},
  author = {Halimeh, Jad C. and Van Damme, Maarten and Zauner-Stauber, Valentin and Vanderstraeten, Laurens},
  journal = {Phys. Rev. Res.},
  volume = {2},
  issue = {3},
  pages = {033111},
  numpages = {8},
  year = {2020},
  month = {Jul},
  publisher = {American Physical Society},
  doi = {10.1103/PhysRevResearch.2.033111},
  url = {https://link.aps.org/doi/10.1103/PhysRevResearch.2.033111}
}

@article{Zhou2018pra,
  title = {Dynamical quantum phase transitions in non-Hermitian lattices},
  author = {Zhou, Longwen and Wang, Qing-hai and Wang, Hailong and Gong, Jiangbin},
  journal = {Phys. Rev. A},
  volume = {98},
  issue = {2},
  pages = {022129},
  numpages = {15},
  year = {2018},
  month = {Aug},
  publisher = {American Physical Society},
  doi = {10.1103/PhysRevA.98.022129},
  url = {https://link.aps.org/doi/10.1103/PhysRevA.98.022129}
}

\end{document}